# DISCRETE-EVENT SIMULATION MODEL FOR DRIVER PERFORMANCE ASSESSMENT: APPLICATION TO THE AUTONOMOUS VEHICLE COCKPIT DESIGN OPTIMIZATION

*Ilya Yuskevich[1,2], Andreas-Makoto Hein[1], Kahina Amokrane-Ferka[2], Abdelkrim Doufene[2], Marija Jankovic[1]*
[1]*CentraleSupélec;* [2]*IRT SystemX*

**Abstract**

The latest advances in the design of vehicles with the adaptive level of automation pose new challenges in the vehicle-driver interaction. Safety requirements underline the need to explore optimal cockpit architectures with regard to driver cognitive and perceptual workload, eyes-off-the-road time and situation awareness. We propose to integrate existing task analysis approaches into system architecture evaluation for the early-stage design optimization. We built the discrete-event simulation tool and applied it within the multi-sensory (sight, sound, touch) cockpit design industrial project.

*model based engineering, design optimisation, ergonomics, human-machine interface, autonomous vehicle*

## 1. Introduction

The accelerating technological progress rapidly changes the landscape of available design options in the automotive domain. Constant innovations in the design and functionality of an increasingly autonomous vehicle drive the evolution of the design of a cockpit. This happens because the increasing autonomy alters the status of the driver (authority of control is shared between driver and car). For example, in December 2017 Waymo LLC (the company is owned by Alphabet) has launched the commercial self-driving car service in Arizona. In their car most of the time driving tasks are performed by machine, though trained driver is still required to take over in case of problems (Korosec, 2019).

An automotive cockpit is an example of a human-machine interface (HMI). HMI can be defined as a technical system that allows human operators to monitor and control the state and behavior of the machine.

In the manual mode the driver is included into the control loop. In other words, he has full control over the car's course (i.e. steering) and acceleration. Accordingly, HMI of such car includes elements of direct control and stabilization – steering wheel, acceleration and brake pedal, tachometer, speedometer, etc.

At the highest level of autonomy the driver is fully excluded from the control loop, i.e. he becomes a passenger. Accordingly, the elements of control over car's trajectory are eliminated. Between these two extreme points there are countless amount of interface design options for so-called semi-automated cars for which authority of control is shared between human and machine (see spectrum of assistance and automation in (Flemisch et al., 2014). Notable examples of such interfaces are the maneuver-based approach (Conduct-by-Wire), and haptic-multimodal approach (Horse-metaphor) (Flemisch et al., 2014).

Another practical semi-automated car design option is the vehicle with the adaptive level of automation. As defined in (Scerbo, 2008) the automation is adaptive when *"the level of automation or the number*

*of systems operating under automation can be modified in real time. In addition, changes in the state of automation can be initiated by either the human or the system"*.

Up to date, infrastructure and technologies are not mature enough to enable the operation of fully autonomous vehicles. Under these conditions, adaptive automation is a good solution compared to fixed task allocation (between machine and human), since it allows to avoid cognitive overloads of a driver, boost situation awareness, or reduce complacency (Parasuraman and Wickens, 2008) depending on the state of a driver or road conditions.

The major shortcoming of the adaptive automation is added complexity to the user interface. Traditional cars' interfaces have only elements of the direct control. Future fully autonomous cars will have only interfaces that communicate to the machine the coordinates of the final destination. At the same time, an interface of a car with the adaptive level of automation maintains elements of direct control, includes some features of self-driving cars and, additionally, is augmented by elements of control over the level of automation (switchers, indicators of level of automation, take-over requests, etc).

Accordingly, interface for a car with the adaptive level of automation may increase workload and eyes-off-the road time, especially at the moments of transition between automation levels. These negative effects can be mitigated by optimizing of the HMI's functional architecture.

This work is intended to review existing methods of workload and eyes-off-the-road time estimation, integrate them into system architecture evaluation process, build the supporting tool and apply it for a new type of systems.

The remainder of the paper is structured as follows. In the next section we present the results of the literature review in the field of workload and eyes-off-the-road time modelling. Then, our proposed discrete-event simulation model is described. After that, we present the results of the case study. In the final section we discuss how the model can be extended and validated in future work.

## 2. Background

### 2.1. Mental workload modeling

Since mid-80s, a lot of research efforts have been devoted to the development of the human operator mental workload modeling tools. Most of these approaches are based on the idea of task analysis, or, more specifically, task network. Task network is a functional decomposition of a human operator's activities down to elementary tasks (Laughery et al., 2000). Then these elementary tasks are annotated with number of attributes (descriptors), e.g. required workload, task duration, triggering event, etc. The total cognitive workload is then calculated either with matrix-based approach (W/INDEX) (North and Riley, 1989) or as the result of a simulation in a discrete-event environment (Aldrich et al., 1989). More recent works in this field propose to employ Petri nets to model human operator strategies, adaptive to the changing environment (Kontogiannis, 2005).

Boy (1998) generalized task network approaches under the name of Cognitive Function Analysis (CFA). This approach responds to the emergence of highly automated and cooperative systems. In CFA, complex cognitive functions may be allocated not only to a human, but also to a machine, which in turn is interacting with other automated machines. Accordingly, CFA is more suitable for the modern context where authority of control is shared between humans and machines.

In the past, all these approaches were driven mainly by increasing automation in military, aerospace or power plants domains. Nowadays, research focus is shifting to the automotive systems due to the progress in the domain of car automation.

### 2.2. Eyes-off-the-road time estimation

Compared to mental workload modeling, eyes-off-the-road time prediction approaches received less attention in the literature. However, this metric is important from safety perspective in the automotive domain. Driver's distraction is the leading defined cause of road accidents according to (Wang et al., 1996). Up to 20% of crashes due to driver distraction are caused by the interaction with interior equipment (car interfaces or cellphone) (Green, 2017). These facts underline the importance of efforts to minimize the potential visual distraction caused by elements of user interface during the design optimization.

In (Wittmann et al., 2006) it was experimentally proven that the location of the onboard display greatly influences the safety of driving, e.g. perception of the information on the head-up display causes shorter distraction compared to instrument cluster or central panel.

The integration of the Keystroke Level Model and occlusion technique was presented in (Pettitt et al., 2007). The goal of this approach is to predict eyes-off-the-road time having a list of driving tasks as an input. The validation of this approach have shown high accuracy with experimental results.

## 2.3. Cognitive simulation models

Approaches based on task analysis treat driver's cognition as a black-box (tasks play role of inputs and outputs). In contrast, cognitive simulation methods approach the modelling of the human internal mental processes. Without the model of the human cognition system it is impossible to evaluate another important safety-related metric – situation awareness.

The most comprehensive model of the human driver performances up to date is COgnitive Simulation MOdel of the DRIVEr (COSMODRIVE) (Bellet et al., 2011). COSMODRIVE is composed of three modules – perception, cognition and action. It models in details all main human driver mental activities. It can be connected to SiVIC virtual road environment platform to provide very detailed input to a perceptual model of the virtual driver.

A workload prediction method based on cognitive architecture for safety critical task simulation (CASCaS) (Feuerstack et al., 2007) was used to build real-time assessment of driver's workload in order to enable adaptive automation (Wortelen et al., 2016).

## 3. Discrete-event simulation model architecture

In this section we describe the architecture of our proposed discrete-event simulation model, task data structure and workload scales.

From systems engineering point of view, the design of the autonomous vehicle cockpit is a challenging task even though the complexity of the physical architecture of this kind of a system is relatively low (compared to, for example, to aerospace systems). The complexity of the design of a cockpit rests on the ambiguity of the functional architecture which builds upon the ambiguity and complexity of the human driver behavior. The latter is complex due to the fact that the number of possible human reactions is large and hardly formalizable, and ambiguous because there are a number of known unknowns such as level of experience, personal attitude, tiredness, mood, etc.

However, we can claim without proof that two functional architectures of a cockpit may be compared in terms of safety without an ambition to predict absolute values of the cockpit's key performances. For example, if one variant of a cockpit constantly transmit a lot of unnecessary information to a driver, the critical information may be eventually missed, which means that other, less distracting variant of a cabin is generally safer.

Our system of interest is a cabin of a car with adaptive levels of automation (Level 0-4). The Level 4 we will call further Automotive Driving (AD) mode. A driver can switch to a higher level of automation at any moment if this level is available due to road conditions (e.g. AD mode may not be available due to the absence of road markings). A higher to a lower level of automation switch is activated by the driver at any moment or by the machine if this higher level of automation is no longer available due to the road conditions.

Special mode called take-over request (TOR60) is activated 60 seconds before the moment when AD is no longer available. During this mode interface sends signals to a driver in order to advise him to put hands on a steering wheel and switch to a lower level of automation. The state machine of the car of interest is shown on figure 1.

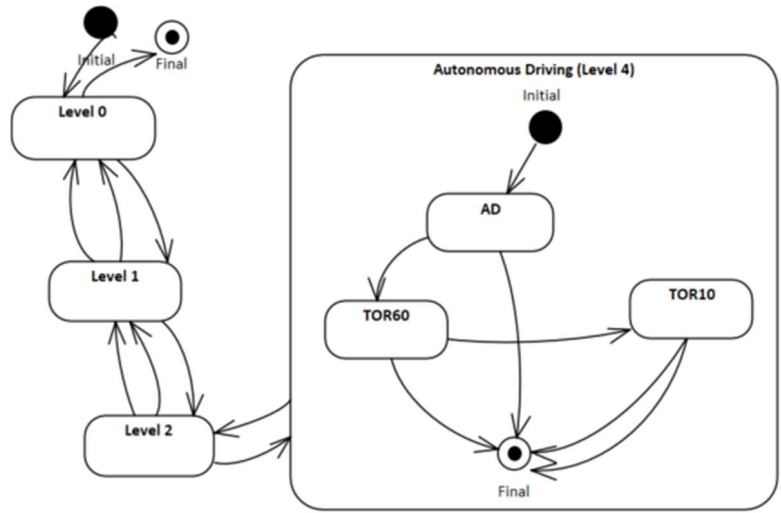

**Figure 1.** The state-machine of the car with adaptive automation

The architecture of the human-machine interaction model is shown on figure 2. It consists of 7 major elements: road conditions, vehicle's state machine and cognitive functions, driver's memory, tasks' schedule and cognitive functions, and task list. Elements are communicating to each other by means of events. For the modeling we use Python discrete-event simulation library SimPy.

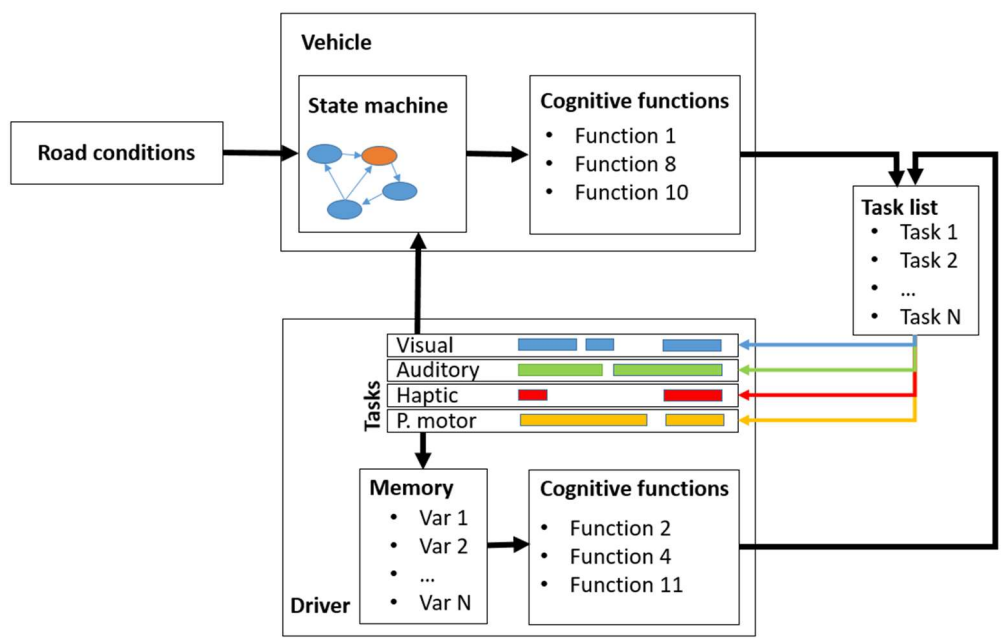

**Figure 2.** Architecture of the discrete-event simulation model of HMI

Event is triggered in one element and influences the processes or state of another element. For example, an event of road conditions change may trigger the transition of the vehicle state (automation level). Cognitive functions are modelled as a random processes generating events in time. For example, "check the speed on average every 20 seconds" is a cognitive function. It triggers events normally distributed in time with mean of 20 seconds.

Tasks are modelled by complex objects which in SimPy referred to as processes. Process is triggered by some event and active within the limited interval of time. When process is active it requires some definite amount of limited resource. If the process is triggered but the resource is taken entirely by other processes, this process is put on the waiting list according to its priority. Processes from the waiting list with higher priority are executed first. After the process execution another event is generated, which triggers another processes, etc.

SimPy object "resource" models the limited capacity of user cognition and perception. According to multiple-resource theory (Wickens, 2002), the human brain can perceive information from two sources efficiently if they require separate attentional resources. There are seven separate attentional resources used in our multi-sensory interface: visual (displays), visual peripheral (ambient lighting), auditory vocal and auditory non-vocal (speakers), haptic hands (steering wheel haptic actuator), haptic seat (seat haptic actuator) and psychomotor. We will use the simplest model of the resource conflict matrix, i.e. two tasks cannot use the same resource concurrently. The sum of active tasks' workloads in our model shall not exceed 10.

We describe each task as an excel row with fixed list of properties (Table 1).

Table 1.  Data structure of a task

| Name of property | Description |
| --- | --- |
| Name | Name of a task |
| Description | Textual description of a task |
| Location | Name of an interface element with which the driver should interact to accomplish a task (e.g. steering wheel, cluster, central console, etc), or, in other words, task-component allocation parameter |
| Cognitive workload descriptor | Textual description of cognitive task complexity |
| Perceptual workload descriptor | Textual description of perceptual task complexity |
| Perception type | Sensorial mode (Visual, Visual peripheral, Auditory Vocal, Auditory non-Vocal, Haptic hands, Haptic seat, Psychomotor) |
| Perceptual workload | Amount of perceptual workload required to accomplish the task (in a relative scale) |
| Cognitive workload | Amount of cognitive workload required to accomplish the task (in a relative scale) |
| Duration | Amount of time needed for task execution |
| Gaze Time | Time needed to change the visual focus from the road to the interface element (only for visual tasks) |
| Total time | Task duration plus gaze time multiplied by two |
| Cognitive function trigger | The name of CPF that triggers the task |
| Awareness parameter | The parameter in user memory that is updated after task execution |
| Triggers | The name of the task that shall be executed right after (if any) |
| Priority | Task relative importance (ordinal scale) |

For example, cognitive function "check the current speed approximately every 10 seconds" generates event that triggers task "check out the speedometer" with "Inspect/Check (numerical)" perceptive descriptor and "Evaluate single aspect" cognitive descriptor. This task requires 0.2 seconds on gaze change from the road to instrument and 1 second to accomplish task. It takes 4.6 points of cognitive resources and 4.0 of visual perception resource. If there are no other active visual tasks, total cognitive workload less than 10 and total perceptual workload less than 10, task is executed during 1.4 seconds.

Accomplished task generates the event which triggers update of variable "current speed" in the object representing driver's memory.

In contrast to tasks initiated by the user (e.g. "check values in the display", "change automation level"), tasks initiated by the machine (e.g. "send vocal message to the user") cannot be put to the waiting list. If such a task is triggered and cognitive or perceptual resources of the driver are taken, then the task is aborted immediately.

We are using workload component scale presented in (Aldrich et al., 1989), to define workload values depending on task complexity (Table 2).

Table 2. Workload component scale derived from (Aldrich et al., 1989)

*Cognitive*

| Descriptor | Workload value |
|---|---|
| Simple association | 1.0 |
| Select alternative | 1.2 |
| Sign/signal recognition | 3.7 |
| Evaluate single aspect | 4.6 |
| Encoding/Decoding/Recall | 5.3 |
| Evaluate several aspects | 6.8 |

*Visual*

| Descriptor | Workload value |
|---|---|
| Detect simple signal | 1.0 |
| Discriminate (Sign) | 3.7 |
| Inspect/Check (numerical) | 4.0 |
| Read (text) | 5.9 |
| Scan/Search/Monitor | 7.0 |

*Auditory*

| Descriptor | Workload value |
|---|---|
| Non-vocal signal recognition | 6.6 |
| Vocal signal recognition | 4.9 |

*Psychomotor*

| Descriptor | Workload value |
|---|---|
| Push the button | 2.2 |
| Switch toggle | 2.2 |
| Continuous adjustive controller | 2.6 |
| Discrete adjustive controller | 5.8 |

*Haptic*

| Descriptor | Workload value |
|---|---|
| Detect simple signal | 1.0 |

Situation awareness is modeled as a correspondence of the values in the driver's memory to the actual state of the car and road conditions. They may not correspond to each other if the task, which is responsible for the driver's memory update is delayed or interrupted due to the overload. This trivial model of situation awareness is far from fidelity of the modern comprehensive models like COSMODRIVE but sufficient for our purposes.

The HMI's functional architecture optimization problem is formulated as follows:

> minimize the number of perceptual and cognitive overloads and eyes-off-the road duration in a given interval of time, by optimizing the list of tasks and changing task-component allocation, wherein the situation awareness shall not go below certain limit.

## 4. Results

The sample of the discrete-event simulation output is shown on figure 3. The top chart shows the road conditions and machine state, the bottom – the full span of the simulation sample (the highlighted area marks the zoomed fragment). The chart in the middle represents cognitive and perceptual workloads of the executed tasks. The pointer between 108 and 110 seconds highlights current task (19_04_26) and shows that at the moment the user has wrong understanding of the road conditions state and autonomous driving mode availability.

We will compare two configurations of the functional architecture – the first is proposed by the company's internal experts in ergonomics (referred further to as base design option) and another one is manually optimized version of this base design option. The values of multiobjective function during this optimization were calculated with our proposed tool.

The examples of the design decisions recommended for the functional architecture optimization are given in table 3.

Table 3. Examples of design decisions

| |
|---|
| Change textual message "AD mode is available" on icon "AD" and reallocate it to Head-up display |
| Change textual message "L1 is activated" on icon "L1" on the Instrument cluster |
| Remove vocal message "Push on button to activate AD mode " |
| Put vocal message "Drive Now" after haptic signal "TOR10" (avoid these signals to appear simultaneously) |
| Put non-vocal message "TOR10" after vocal message "Drive Now" (avoid these signals to appear simultaneously) |

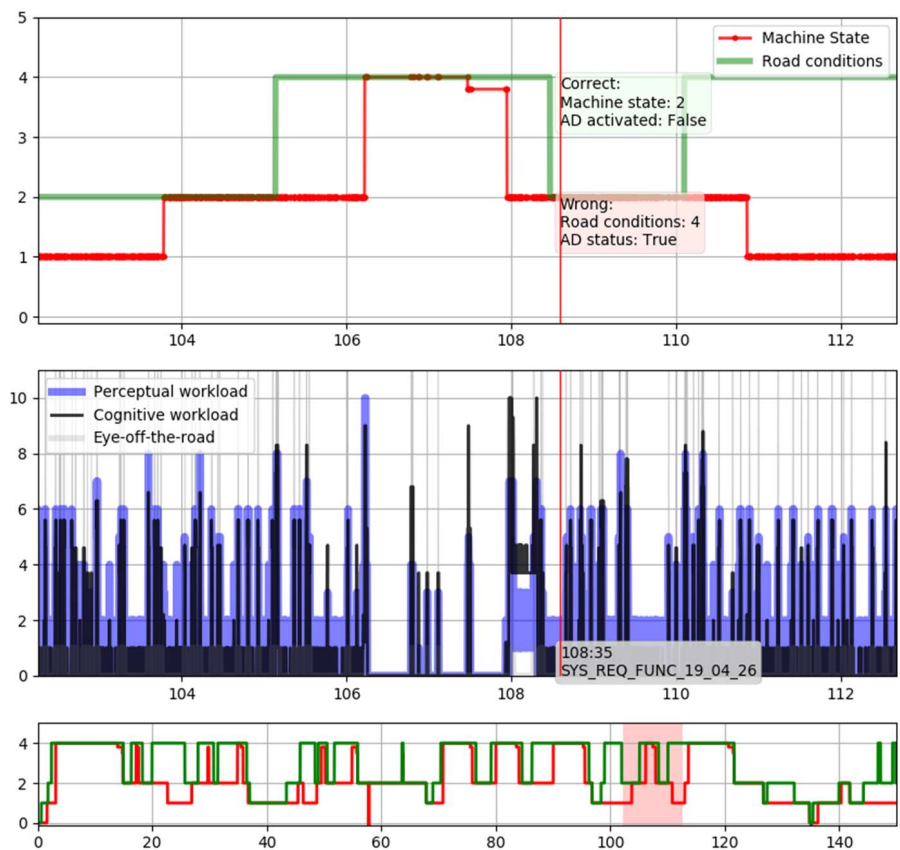

Figure 3. Timescale output of the discrete-event simulation

Discrete-event simulation environment uses pseudo-random generators to trigger events. Hence, to compare two functional architecture configurations we should execute a number of trials. The simulation time of each trial is 1000 minutes of virtual driving (computational time is around 30 seconds for each trial). Figures 4 and 5 show the statistics of the task execution in random trials for two architectures.

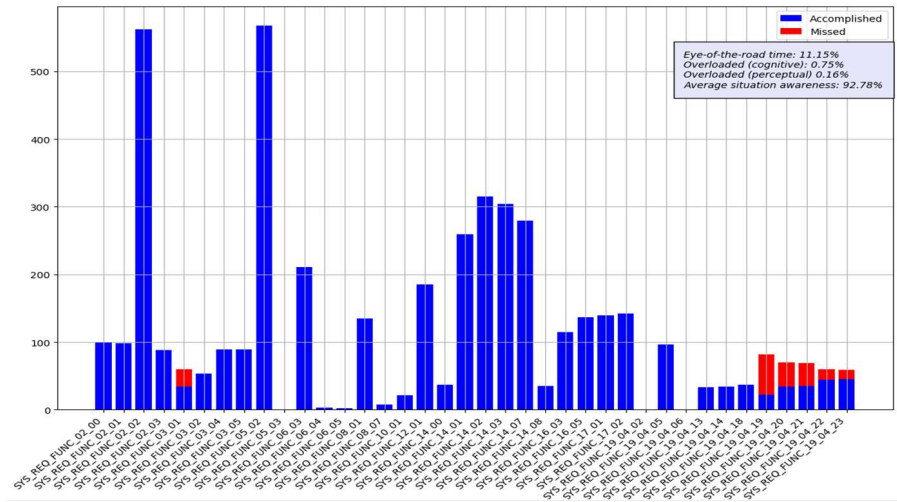

Figure 4. Base design option simulation statistics

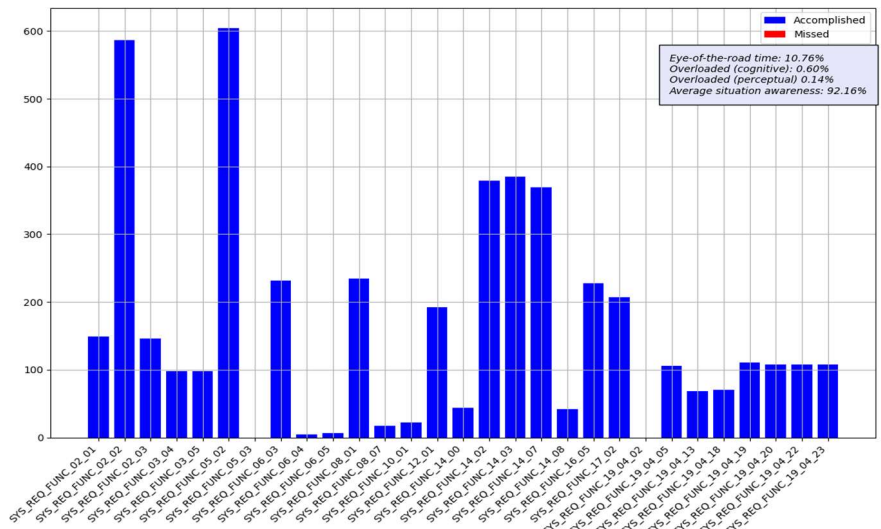

Figure 5. Optimized design option simulation statistics

On figure 6 the results of 40 simulations (20 for the base and 20 for the optimized configuration) are shown in two-dimensional space.

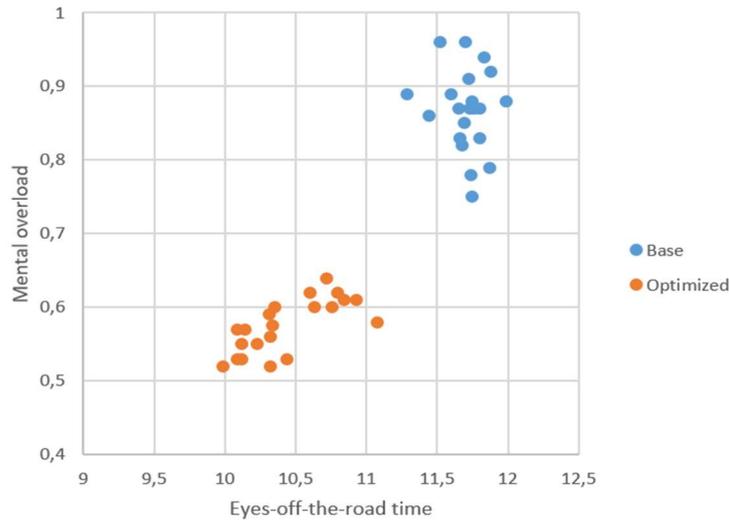

**Figure 6.** The results of discrete-event simulations of the base and optimized design option

**Table 4.** Averaged results of two design options

| Indicator | Base | Optimized |
|---|---|---|
| Median eyes-off-the-road time, % of total time | 11.7 | 10.3 |
| Median cognitive overload, % of total time | 0.87 | 0.57 |
| Median perceptual overload, % of total time | 0.19 | 0.13 |
| Median situation awareness, average % | 92.1 | 92.3 |

# 5. Conclusions and future work

In this paper we have shown how existing task analysis approaches can be applied to the design of a modern vehicles with the adaptive level of automation. Our model designed to enable fast early-stage functional architecture optimization with respect to widely used safety metrics: cognitive and perceptual overload, eyes-off-the-road time and situation awareness.

We applied our model to the real industrial project and obtained a list of recommendations to improve the current conceptual design. Still, we can point out several directions for future work in order to improve accuracy and validity of our model:

- validation of the obtained workload and eyes-off-the-road results on the simulator with real humans;
- integration to our model more sophisticated and accurate conflict matrix (measure of a conflict between pair of perceptual resources);
- verification of the models of human driver's cognitive functions (e.g. how often driver checks the state of the machine?);
- functional architecture optimization subject to different human profiles (experienced/novice driver, open-minded/conservative);
- integration of models of human performance degradation (fatigue and drowsiness) and improvement (learning-curve) with time.